\documentclass[apl,amsmath,amssymb,reprint, twocolumn]{revtex4}

\usepackage{graphicx}
\makeatletter
\renewcommand{\section}{\@startsection{section}{1}{0mm}
  {-\baselineskip}{0.5\baselineskip}{\bf\leftline}}

\makeatother

\begin{document}

\title{Cryogenic LED pixel-to-frequency mapper for kinetic inductance detector arrays}

\author{X. Liu$^1$}
\author{W. Guo$^1$}
\author{Y. Wang$^1$\footnote{Electronic mail: qubit@home.swjtu.edu.cn}}
\author{L. F. Wei$^1$}
\author{C. M. Mckenney$^2$}
\author{B. Dober$^2$}
\author{T. Billings$^3$}
\author{J. Hubmayr$^2$}
\author{L. S. Ferreira$^2$}
\author{M. R. Vissers$^2$}
\author{J. Gao$^{2}$\footnote{Electronic mail: jgao@boulder.nist.gov}}

\affiliation{
1) Quantum Optoelectronics Laboratory, School of Physical Science and Technology, Southwest Jiaotong University, Chengdu, Sichuan 610031, China\\
2) National Institute of Standards and Technology, Boulder, CO 80305, USA\footnote{Contribution of the U.S. government, not subject to copyright}\\
3) University of Pennsylvania, Philadelphia, PA 19104, USA}
\date{\today}

\begin{abstract}

We present a cryogenic wafer mapper based on light emitting diodes (LEDs) for spatial mapping of a large microwave kinetic inductance detector (MKID) array. In this scheme, an array of LEDs, addressed by DC wires and collimated through horns onto the detectors, is mounted in front of the detector wafer. By illuminating each LED individually and sweeping the frequency response of all the resonators, we can unambiguously correspond a detector pixel to its measured resonance frequency. We have demonstrated mapping a 76.2~mm $90$-pixel MKID array using a mapper containing $126$ LEDs with $16$ DC bias wires. With the frequency to pixel-position correspondence data obtained by the LED mapper, we have found a radially position-dependent frequency non-uniformity $\lesssim 1.6\%$ over the 76.2~mm wafer. Our LED wafer mapper has no moving parts and is easy to implement. It may find broad applications in superconducting detector and quantum computing/information experiments.

\end{abstract}

\maketitle

\section{Introduction}
Microwave kinetic inductance detectors (MKIDs)~\cite{Day,Doyle,Baselmans} are superconducting pair breaking detectors based on high-quality (high-$Q$) superconducting resonators~\cite{zmuidzinas,yiwen}. MKIDs have received great attention in astronomy~\cite{Eyken,Sayers,Monfardini} and other sensitive detection fields~\cite{Rowe,Jiansong} because they are simple to fabricate and easy to multiplex into large arrays. A MKID array with thousands of pixels can be fabricated with one or a few standard photo-lithography steps and read out with a pair of coaxial cables into the cryostat.

For a large MKID array, with hundreds or thousands of pixels, an important task is to definitively correspond each physical pixel to its measured resonance frequency. In certain applications where the resonance frequency spacing is large, the resonators can be easily identified by their designed frequencies. In most cases, however, resonators are packed as close as possible in frequency space to maximize the multiplexing factor. The measured resonance frequencies are shifted unpredictably from their designed values due to imperfections in the fabrication process, such as variations in superconducting transition temperature ($T_c$), film thickness, and over-etching depth. For these reasons, mapping the measured resonance frequencies to the physical pixel locations on a large MKID array is usually a challenging task.

For example, we are developing feed-horn coupled dual-polarization sensitive MKID~\cite{Hubmayr} arrays made from titanium-nitride/titanium (TiN/Ti) multilayer films with target $T_{c}\approx 1.4$~K for the BLAST-TNG experiment~\cite{Galitzki}. We have made a 90 pixel hexagonal close-packed TiN MKID array on a 76.2~mm intrinsic Si wafer to study the wafer uniformity (Fig.~1(a)). Each pixel is identical to the BLAST-TNG pixel design~\cite{Dober} in the 500 $\mu$m band, which consists of orthogonal TiN absorbers~\cite{Vissers, Vissers2} attached to a pair of lumped-element resonators (Fig.~1(b)). The 5~mm pixel-to-pixel spacing is also the same as in the BLAST-TNG 500 $\mu$m MKID array. The measured $S_{21}$ transmission of the sub-array including all the X-pol resonators (Fig. 1(c)) shows $90$ resonance dips in a narrow frequency range spanned from $530$ MHz to $590$ MHz. The number of resonances and the frequency range match the designed pixel number and the target frequency band. Although the resonators are designed to have evenly spaced frequencies by systematically varying the size of interdigitated capacitor (IDC), the measured resonance frequencies clearly deviate from such a regular pattern, making it impossible to unambiguously correspond a physical pixel to its actual frequency from the transmission data. A common approach to identification is cryogenic beam mapping, which typically requires a vacuum-sealed window, IR filters to control radiative load, and complicated coupling optics~\cite{Eyken,Sayers}.

In this paper, we demonstrate a cryogenic wafer mapper based on light emitting diodes (LEDs), which is easy to implement and has low cost, low power dissipation and no moving parts at cryogenic temperature. We further demonstrate using this tool to determine the wafer uniformity.

\section{LED array}
\begin{figure}[ht]
\includegraphics{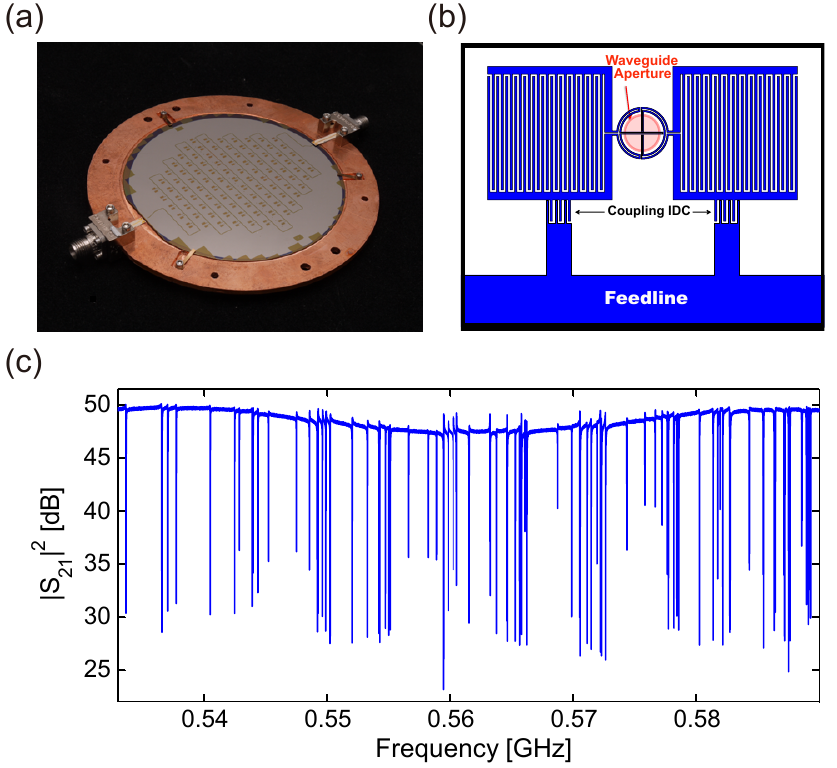}
\caption{ (a) Photo of the 76.2~mm MKID array mounted in a circular sample holder with input/output SMA connectors. (b) Schematic drawing (not to scale) of the single pixel design showing orthogonal X- and Y-polarization sensitive TiN absorbers attached to a pair of lumped-element MKIDs. The waveguide aperture, which illuminates the inductive parts, is depicted by the red shaded circular region. (c) $S_{21}$ measured at 40 mK using a vector network analyzer. $90$ resonances appear in the designed $\sim$ $60$ MHz frequency span around $0.56$ GHz, corresponding to the 90 X-pol resonators. The 90 Y-pol resonators are designed to be in a different frequency band.}
\end{figure}

To make a wafer mapper for MKID arrays, we started by looking for a small-sized and energy-efficient element that can be close-packed, individually addressed and can perturb a single detector pixel with an identifiable response distinguishing it from other unperturbed MKID pixels. At the same time this element should be able to work at low temperature and with easy connection and control. Inspired by Ref.~\cite{Forgues}, we find LEDs are ideal elements for this task. In this reference, a number of LED models are verified to emit light at cryogenic temperature down to 4~K. As for our wafer mapper, we require LEDs that can emit weak optical power on the order of 100~nW at a few millikelvin (mK) to induce a measurable frequency shift in the resonator. In our work, we have verified many LED models that satisfy our requirement at a few mK and cause negligible thermal dissipation. In addition, LEDs have a variety of sizes and shapes, such as round-shaped LED with 3~mm or 5~mm diameters and surface mount LEDs as small as 1~mm by 0.5~mm. It is not difficult to choose a LED size that fits the spacing of the pixels. It is also not difficult to collimate visible light onto individual pixels through horns or lenses. Last, common LEDs are very inexpensive commercial products (less than $\$1$ per LED) and the total cost to build the LED mapper with 126 LEDs described in this letter is less than $\$500$.

We designed a printed circuit board (PCB) and populated 126 round LEDs (3~mm in diameter) onto this PCB. The LED array has the same hexagonal packing scheme as the pixels on the Si wafer (see Fig.~2(a)). To bias and address the 126 LEDs with a minimal number of DC wires, we divide the hexagon into two regions each containing 63 LEDs. We then group the 63 LEDs in each region into 9 rows and 7 columns, which requires $9+7=16$ addressing wires. We further take advantage of the polarity of LEDs to multiplex the wires going to the two regions, which doubles the number of LEDs that can be addressed individually (see supplementary material for more details on the wiring scheme of the 126-LED MKID wafer mapper). In the end, we used a total of 16 DC wires and wired all the 126 LEDs to a 16-pin connector on the PCB. In addition, we designed an aluminum lid (Fig.~2(b)) with a hexagonal array of drilled horns (1~mm in diameter), which serve as collimators. When assembled together at room temperature with a few locking screws, the center positions of the LEDs on the PCB and the horn collimators on the lid are aligned to the center positions of the inductive absorber strip (Fig.~2(c)). We used a microscope to check the alignment by directly observing whether the ``cross'' formed by the orthogonal absorbers of each pixel lies in the center of the horn aperture. The alignment is verified to be better than 100~$\mu$m on average. When one of the LEDs is turned on,  the light will be shining onto one pixel. By individually turning on each LED, we can measure the frequency shifts of all the resonators. The resonator showing the largest response should be the pixel located below the LED. In this way, we can correspond each physical pixel with its measured resonance frequency.

\begin{figure}[ht]
\includegraphics{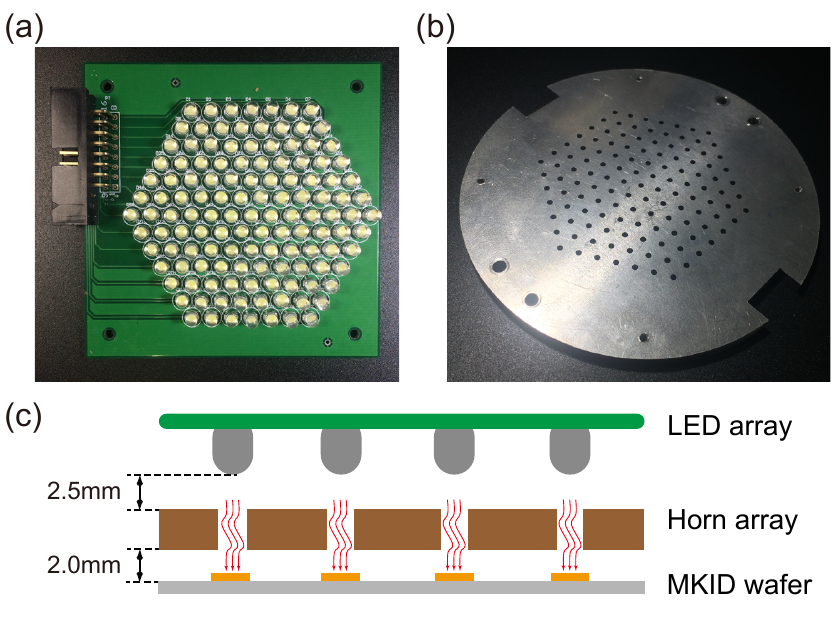}
\caption{(a) A total number of $127$ LEDs are mounted on the PCB. The hexagonal LED array matches the target pixel array shown in Fig. 1(a). The LEDs, except for the center one, are wired to a black 16-pin connector on the top left corner. So there are 126 active LEDs that can be addressed. (b) Horn array drilled on the aluminum lid as collimators. (c) The horn array matches with both the LED array on the PCB and the MKID array on the wafer so that the light emitted from a single LED is collimated onto the corresponding pixel on the wafer.}
\end{figure}

\section{Experiments}
The assembly, with the MKID wafer, horn array and LED array aligned together, is mounted on the mixing chamber (MC) stage in a dilution refrigerator (DR) and cooled down to a base temperature of $40$ mK. We can conveniently apply voltage/current to an individual LED using a breakout box and voltage/current source at room temperature. In the MKID array design, we separated the X-pol resonators and Y-pol resonators into two different frequency bands. Here we only study the $90$ X-pol resonators since they are distributed in a much narrower band and it is more challenging to distinguish them due to the smaller frequency spacing. We first measured the transmission $S_{21}$ of the MKID array in the dark by using a vector network analyzer (VNA) sweeping in the designed frequency band centered at $\sim 0.56$ GHz. As shown in Fig. 1(c), $90$ resonance dips are clearly observed and all the X-pol resonators are present, suggesting the yield is very high ($100\%$ for this fabrication). We index these resonances as Res. $1$, Res. $2$, $\cdots$, Res. $90$ respectively in ascending frequency order in our following discussion.

In our initial experiment, $5$ mm round LEDs (OPTEK OVLEW1CB9) were used since they were reported to work down to $3$ K according to Ref.~\cite{Forgues}. Later we found that $3$ mm round LEDs (Risym F3) are also able to work for our wafer mapper at ultra-low temperature down to $40$ mK. They are used for our final experiment reported in this letter because their smaller size makes the alignment easier. We turn on a specific LED by applying a current of $0.08$ $\mu$A (or a voltage of 3.58~V) through the corresponding address lines. When a single LED is turned on, we sweep all the resonances again and measure the frequency shifts compared to the dark sweep data. All the resonance frequencies are determined by the fitting procedures outlined in Ref.~\cite{Gao}. The LED induced fractional frequency shift $\delta f_r/f_r$ vs. resonator index is shown in Fig. 3(a). Here, $\delta f_r/f_r = (f_{r0}-f_{r})/f_{r0}$, where $f_{r0}$ is the resonance frequency in the dark and $f_{r}$ is the resonance frequency under illumination. One can see that the frequency of one resonator (Res. $52$) shifts significantly more than other resonators, suggesting Res. $52$ is the resonator directly under the illuminated LED. Note that the responses from the other resonators are nonzero, which may be attributed to the light (reflected from the metal surface, Si surface or box walls) inside the device box leaking to other pixels, especially those next to the illuminated resonator. In future experiment, we plan to further reduce the leakage light by applying light absorbing material inside the box. Nevertheless, the pixel responses from unilluminated pixels are at least an order of magnitude lower than the illuminated pixel. Therefore we can identify the physical pixel corresponding to Res. $52$ with a very high confidence level.

\begin{figure}[ht]
\includegraphics{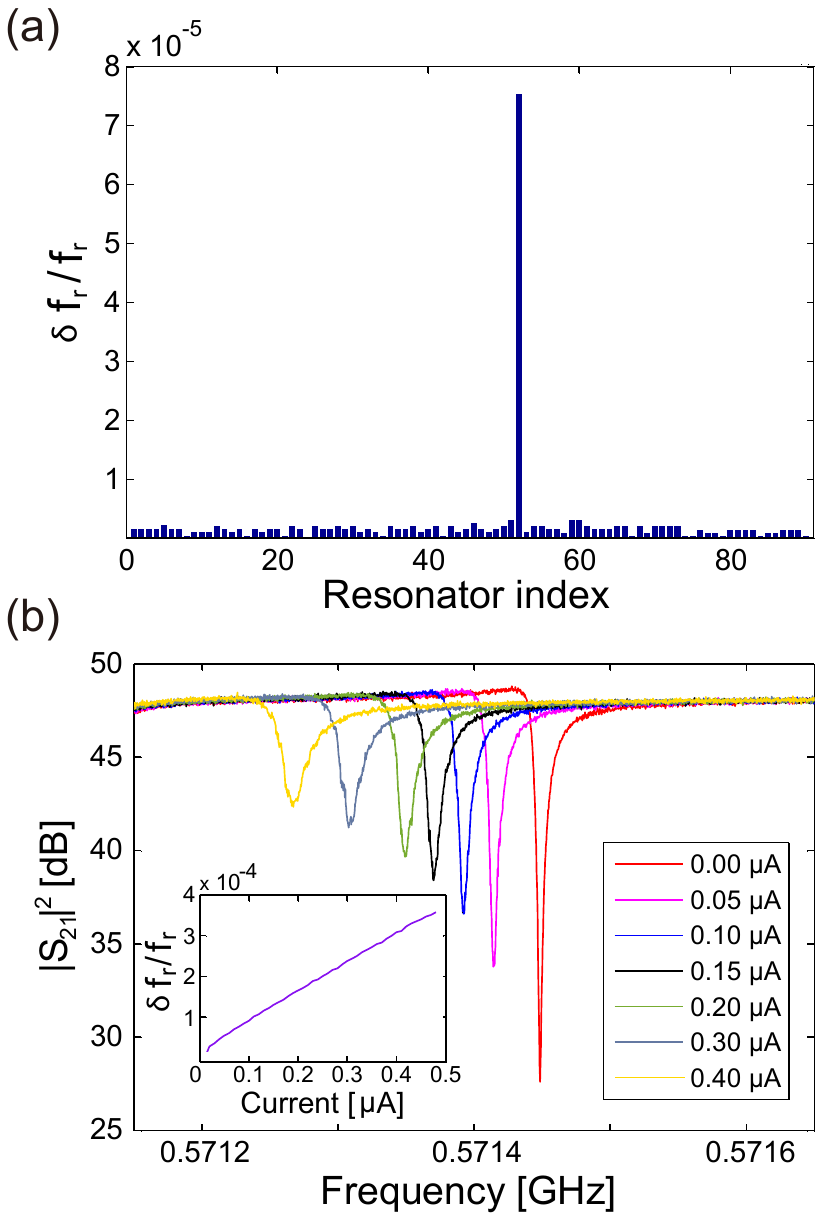}
\caption{(a): The fractional frequency shifts vs. resonator index for all 90 X-pol resonators when a single LED is illuminated. Res. $52$ shows the largest response indicating that Res. $52$ corresponds to the pixel beneath the illuminated LED. (b): $S_{21}$ of Res. $52$ as a function of LED bias current. The inset shows the fractional frequency shift is approximately linear with the LED bias current.}
\end{figure}

Next we measured the frequency responses of Res. $52$ under different LED bias currents. The result is plotted in Fig. 3(b). As we increase the bias current, both the resonance frequency and the quality factor decrease due to photon-induced Copper-pair breaking. As shown in the inset, the fractional frequency shift is almost linear with the LED bias current $I$. Our result implies that the resonators have a linear frequency response to the optical power, i.e., the fractional frequency shift $\delta f_r/f_r \propto P_{opt}$, considering $P_{opt}\propto I\Delta$. Here, $P_{opt}$ is the applied optical power and $\Delta$ the band gap of the LED. This linear optical response for TiN MKID has been reported earlier in sub-millimeter wave measurement and the underlying physics is now an active area of research~\cite{Hubmayr}. It is interesting to observe the linear response of TiN MKID to optical photons whose energy is 3 orders of magnitude higher than mm-wave photons. This observation provides valuable information for the study of disordered superconductors.

At the bias condition $I = 0.08$ $\mu$A and $V = 3.58$~V used for Fig.~3(a), we estimate the power dissipated on the mixing chamber plate to be less than $0.29$ $\mu$W, far less than the cooling power of our DR. $V$= 3.58~V is both the voltage output of the power supply and the actual voltage dropped on the LED, because the bias wire resistance (a few tens of $\Omega$) is orders of magnitude smaller than the resistance of the LED (a few M$\Omega$) at 40~mK. Furthermore, we pulsed the LED to generate a light pulse of 200~ns in width and the recovery time of the resonance response of all the resonators is measured to be less than $10$ $\mu$s, suggesting that the frequency shift of the illuminated resonator is dominated by optical response instead of thermal effect (the time scale is on the order of a few seconds) and that the response from other unilluminated resonators is mainly due to leakage light instead of the whole wafer warming up.

\begin{figure}[htb]
\includegraphics{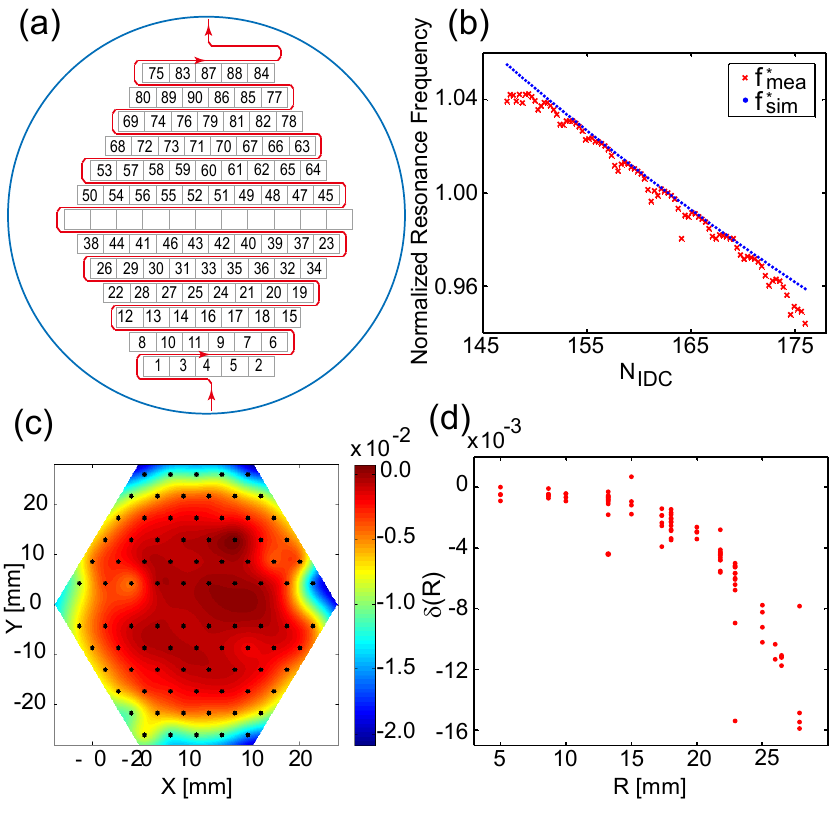}
\caption{Wafer uniformity information extracted by using the LED mapper. (a) The map of the physical pixels in the order of their measured frequencies. The blue circle represents a 76.2~mm MKID wafer. The red path line indicates the feedline, and the arrow shows the direction of increasing frequency in design. The pixels on the center row are diced out for a separate measurement (not discussed in this letter). They are left as blanks in the wafer map. (b) The normalized design (simulated) resonance frequency $f^*_{sim}$ and measured resonance frequency $f^*_{mea}$ vs. the number of IDC fingers $N_\mathrm{IDC}$. Both lines are normalized by their values for the pixel with $N_\mathrm{IDC} = 160.8$, i.e., $f^*_\mathrm{mea} = f_\mathrm{mea}/f_\mathrm{mea}(N_\mathrm{IDC} = 160.8)$ and $f^*_\mathrm{sim} = f_\mathrm{sim}/f_\mathrm{sim}(N_\mathrm{IDC} = 160.8)$.  $f_{sim}$ was derived from Sonnet (a commercial EM Software) simulation using the method described in Ref.~\cite{Wisbey} by assuming a kinetic inductance of $L_\mathrm{ki} = 20$ pH for the TiN/Ti multilayer film and a dielectric constant of $11.7$ for the Si substrate. (c) A color-coded surface plot visualizing the position-dependent fractional frequency deviation $\delta$ on the 76.2~mm wafer. The black dots indicate the center positions of the measured pixels, whose values are linearly interpolated to generate the 2D plot over the entire area. (d) $\delta$ of the pixels vs. their distance to the wafer center $R$ clearly shows a radial dependence.}
\end{figure}

\section{Wafer non-uniformity}
By turning on each LED individually, we mapped out all the physical pixels corresponding to their measured resonance frequencies, as shown in Fig. 4(a). The frequency of the MKID array was initially designed to monotonically increase along the winding direction of the meandering feedline (the red path line in Fig. 4(a)) by gradually decreasing the total number of IDC fingers $N_\mathrm{IDC}$. The measured frequencies roughly follow this trend but frequency re-shuffling occurs on a few adjacent resonators. With the measured frequency to pixel correspondence data, we can further study the wafer uniformity. We first compare the measured resonance frequency $f^*_\mathrm{mea}$ with the designed resonance frequency $f^*_\mathrm{sim}$ for each pixel, which is plotted by the blue and red curves in Fig.~4(b), respectively. Note that both lines have been normalized by the resonance frequency of the pixel with $N_\mathrm{IDC} = 160.8$ which is located near the center of the wafer. In Fig.~4(b), we see a periodic deviation between the measured and designed frequency, suggesting wafer non-uniformity. We define a position-dependent fractional frequency deviation $\delta =(f^*_\mathrm{mea} -f^*_\mathrm{sim})/f^*_\mathrm{sim}$. The maximum deviation $|\delta|_\mathrm{max}$ is a figure of merit that measures the frequency uniformity over the entire wafer. Small $|\delta|_\mathrm{max}$ indicates better wafer uniformity. A color-coded surface plot visualizing the position-dependent fractional frequency deviation $\delta$ over the hexagonal area covered by the pixels is shown in Fig. 4(c). It exhibits a radially decreasing pattern of $\delta$, and this trend is more evident in Fig.~4(d) which plots $\delta$ as a function of the radial position $R$. From Fig.~4(c) and (d) we conclude that the frequency non-uniformity of our wafer is $|\delta|_\textrm{max} \lesssim 1.6\%$ and the largest deviation occurs on the edge of the wafer.

There are several factors that may contribute to the radial frequency non-uniformity revealed in Fig.~4. It is important to determine whether this non-uniformity is associated with capacitance variation or inductance variation across the wafer. For example, the etch chemistry used to pattern the TiN IDC will over-etch into the Si substrate and a non-uniform over-etch depth distribution may lead to capacitance variation across the wafer. To test this hypothesis, we further measured the over-etch depths for several resonators across the wafer by using a profilometer and found that the over-etch depths range from 136~nm to 141~nm. Electromagnetic simulation using Sonnet software shows that the fractional capacitance change corresponding to the minimum and maximum over-etch depths of our IDC with 2~$\mu$m finger/gap width is $\Delta C/C \approx$ $0.13\%$. This translates into a maximum of $0.06\%$ fractional frequency deviation, which is too small to explain the $1.6\%$ frequency non-uniformity. This suggests that the over-etch effect is not the major factor to produce the measured frequency non-uniformity. On the other hand, cross-wafer variation of film $T_c$, normal resistivity $\rho_n$ and thickness $t$ may lead to inductance variation through the relation $L_\mathrm{ki} = \frac{\hbar R_\mathrm{sn}}{\pi \Delta} \propto \frac{\rho_\mathrm{n}}{t T_c}$\cite{PropTiN}, where $R_\mathrm{sn}$ is the normal sheet resistance and $\Delta\approx 1.76k_{B}T_c$ is the superconducting gap. Previous measurements of multilayer TiN films show small non-uniformity and decreasing $T_c$ with radius~\cite{Vissers2}. A lower $T_c$ results in a larger kinetic inductance $L_\mathrm{ki}$ and a lower resonance frequency. The measured $1.6\%$ frequency non-uniformity suggests a $3.2\%$ $T_c$ non-uniformity, which agrees well with our previous $T_c$ measurement~\cite{Vissers2}. We think this is the dominant contribution to the measured frequency non-uniformity.

\section{Discussions and Conclusions}
We have also tested a number of other LED models, including both circular and surface mount types. All the LEDs in our test are able to illuminate and induce a frequency shift in the MKID, which provides a wider selection for different pixel packing densities. The LEDs we verified to work for our purpose at low temperature down to $40$ mK are summarized in Table~$1$. It should be noted that we are operating the LEDs in a weakly turn-on condition, different from a few other cryogenic applications of LEDs~\cite{Forgues}. In our case, biased at the same forward voltage of $3.58$ V, the LED forward current decreases from 59~mA at room temperature to 0.08~$\mu$A at 40~mK and the emitted optical power reduces from $>$ 100~mW to $\sim100$~nW (assuming $50\%$ optical conversion efficiency), mainly due to carrier freeze-out ~\cite{Fred}. Because our detectors are designed to be sensitive to pW of mm-wave loading power~\cite{Hubmayr}, 100 nW of optical power is sufficient to induce a significant shift in the resonator while the power is still small enough for the fridge to handle. We speculate that many other LED types and models other than the ones listed in Table~$1$ may work for our pixel-to-frequency mapper at millikelvin temperature.

\begin{table}
\newcommand{\tabincell}[2]{\begin{tabular}{@{}#1@{}}#2\end{tabular}}
\centering
\begin{tabular}{|c|c|c|}\hline
LED Model & Type \\\hline
OPTEK OVLEW1CB9 & \tabincell{c}{round, $\Phi$ 5 mm, white}   \\\hline
Risym F3 & \tabincell{c}{round, $\Phi$ 3 mm, white}   \\\hline
\tabincell{c}{HSMD-C190} & \tabincell{c}{SMD0603, ~1.6 mm * 0.8 mm, orange~}  \\\hline
\tabincell{c}{LTST-C193TBKT-5A} & \tabincell{c}{SMD0603, ~1.6 mm * 0.8 mm, blue~}  \\\hline
\end{tabular}
  \caption{LED models verified to work for our mapper at 40 mK.}
\end{table}

In conclusion, we have experimentally demonstrated a cryogenic LED wafer mapper that can spatially map a MKID array at millikelvin temperatures. It provides a simple and effective way to correspond a physical pixel on the wafer to its measured resonance frequency. As a demonstration of this powerful tool, we have studied the wafer uniformity and found a radially position-dependent frequency non-uniformity $\lesssim 1.6\%$ over the 76.2~mm wafer using the frequency-to-pixel correspondence information obtained by the LED mapper. Made from LEDs and PCB, our wafer mapper has very low cost and is very easy to implement, requiring minimal number of DC wires and dissipating only sub-microwatts of power. In an optimized wiring scheme that takes advantage of the LED polarity, one can use $m$ wires to address a maximum number of $P(m,2) = m(m+1)$ LEDs, where P denotes permutation. For example, an array of $1000$ pixels can be mapped out by a LED mapper with only 33 bias wires. By using LEDs of smaller footprint, such as surface mount LED SMD0402 ($1.0$ mm * $0.5$ mm), our mapper can be applied to more densely packed arrays. The LED mapper may find broad applications in which an array of elements (such as superconducting resonators, detectors, and qubits) sensitive to light are multiplexed in the frequency domain.

\section{Supplementary Material}
More details on the wiring scheme of the 126-LED MKID wafer mapper are provided.
\begin{large}
\section*{Acknowledgement}
\end{large}

The MKID devices were fabricated in the NIST-Boulder microfabrication facility. This work was supported in part by the National Natural Science Foundation of China (Grant Nos. 61301031, U1330201).

\end{document}